\documentclass[natbib,twocolumn,twoside]{svmultiag}
\usepackage{graphicx}

\title*{On the selection of prospective sources for ICRF extension}
\titlerunning{On the source selection for ICRF extension}

\author{Z.~Malkin}
\authorrunning{Z. Malkin}

\institute{Zinovy Malkin \at Pulkovo Observatory, Pulkovskoe Sh. 65, St. Petersburg 196140, Russia}

\begin{document}
\maketitle
\thispagestyle{empty}
\pagestyle{empty}

\abstract{
Despite continuous increasing of the number of ICRF sources, their sky coverage is still not satisfactory.
The goal of this study is to discuss some new considerations for extending the ICRF source list.
Statistical analysis of the ICRF catalog allows us to identify less populated sky regions where
new ICRF sources or additional observations of the current ICRF sources are most desirable to improve
both the uniformity of the source distribution and the uniformity of the distribution of the position errors.
It is also desirable to include more sources with high redshift in the ICRF list.
These sources may be of interest for astrophysics.
To select prospective new ICRF sources, the OCARS catalog is used.
The number of sources in OCARS is about three times greater than in the ICRF3, which gives us an opportunity
to select new ICRF sources that have already be tested and detected in astrometric and geodetic VLBI experiments.
}


\section{Introduction}
\label{SEC:introduction}

The International Celestial Reference Frame (ICRF) is the standard of the celestial reference frame
recommended by the International Astronomical Union (IAU) for scientific researches and practical
applications in various fields.
The third ICRF realization, ICRF3 \citep{Charlot2020} approved by the IAU in 2018 is currently in use.
It contains positions of 4588 sources included in three catalogs ICRF3-SX, ICRF3-XKa, and ICRF3-K containing
source positions determined from observations at three radio frequency bands.
The ICRF3-SX catalog containing of 4536 sources observed at the $S$/$X$ band is most complete and it is
currently used to link ICRF and $Gaia$-CRF catalogs.

Although the number of sources in the ICRF catalog is continuously increased with time and the accuracy
of their positions is continuously improved, the sky coverage by the ICRF sources, its density
and uniformity, as well as the sky distribution of the source position errors, are still not satisfactory.
For this reason, several institutions are considering new observing programs aimed at improving the ICRF.
Statistical analysis of the ICRF catalog allows us to identify less populated sky regions where
new ICRF sources or additional observations of the current ICRF sources are most desirable to improve
both the uniformity of the source distribution and the uniformity of the distribution of the position errors.
One more consideration for extending the ICRF list is including of sources with high redshift.
These sources may be of interest for astrophysics.

The main goal of this work is to discuss new approaches for improving the source list for further ICRF extensions.
Especial attention will be given to formalization and quantification of some criteria (mostly known) for
the selection of new sources for future ICRF extensions, as well as identification of the ICRF sources that
need more observations to provide more uniform sky distribution of the position errors.

To select prospective new ICRF sources, the catalog OCARS (Optical Characteristics of Astrometric Radio Sources)
\citep{Malkin2018}) is used.
The current version of the OCARS catalog includes more than 13 thousand sources which is about three times greater
than the ICRF3 catalog.
The OCARS non-ICRF sources mostly have lower position accuracy compared to ICRF, but these sources
have already been tested and detected in VLBI experiments.
This gives us an opportunity to select new ICRF sources from the OCARS list without resource consuming detection tests,
which are necessary if new sources are pre-selected from general radio surveys.

\section{Analysis of possibilities for improving ICRF source list}

This study is based on the analysis of the sky distribution of the ICRF3 and $Gaia$ catalogs, and position
errors of the ICRF3 sources.
To quantify the uniformity of the source sky distribution a method of pixelization of the spherical surface SREAG
(Spherical Rectangular Equal-Area Grid) \citep{Malkin2019,Malkin2020}) is used.
It provides rectangular equal area cells with wide range of resolutions from $\sim$45$^{\circ}$ to $\sim$16$''$
(for 31-bit integer arithmetic).
Fortran routines to perform basic operations with the SREAG pixelization are
provided\footnote{http://www.gaoran.ru/english/as/ac\_vlbi/}.
Table.~\ref{tab:adjusting_sreag} shows basic parameters of several grids that looks most relevant for ICRF studies.
In this paper, the grid with $N_{ring}$=10 (128 cells) has been used for plots of the sky distribution of the ICRF3-SX, $Gaia$,
and OCARS sources, and the grid with $N_{ring}$=18 (412 cells) has been used for plots of the sky distribution of the ICRF3
defining sources. 

\begin{table}
\centering
\caption{SREAG grid parameters as a function of the number of rings $N_{ring}$.}
\label{tab:adjusting_sreag}
\tabcolsep 10pt
\begin{tabular}{ccc}
\hline
$N_{ring}$ & $N_{cell}$ & Cell area \\
&& [sq. deg] \\
\hline
    4    &  20 & 2063 \\
    6    &  46 &  897 \\
    8    &  82 &  503 \\
   10    & 128 &  322 \\
   12    & 184 &  224 \\
   14    & 250 &  165 \\
   16    & 326 &  127 \\
   18    & 412 &  100 \\
   20    & 508 &   81 \\
\hline
\end{tabular}
\vspace{4ex}
\end{table}

\begin{figure}
\centering
\includegraphics[width=0.48\textwidth]{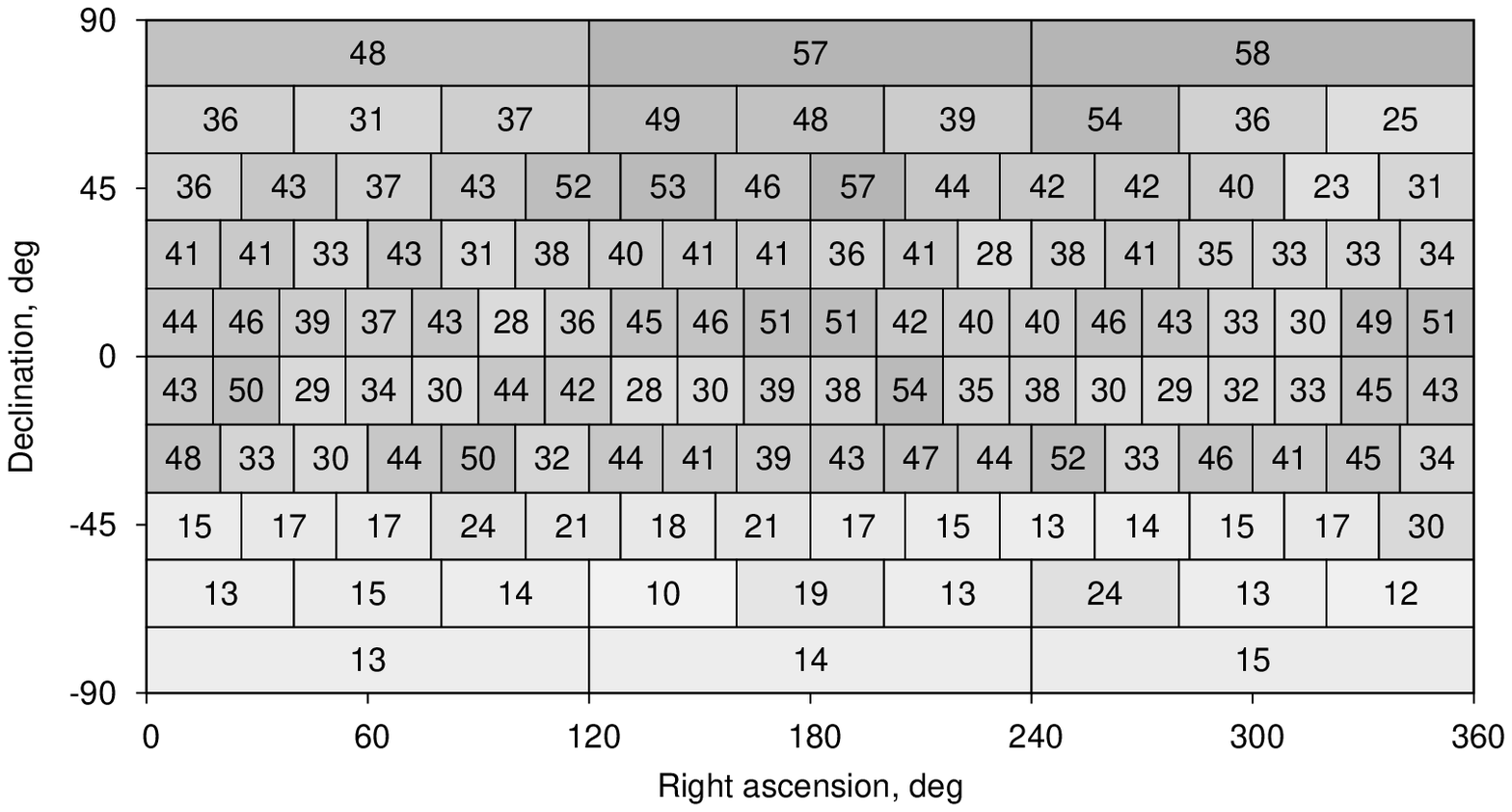}
\includegraphics[width=0.48\textwidth]{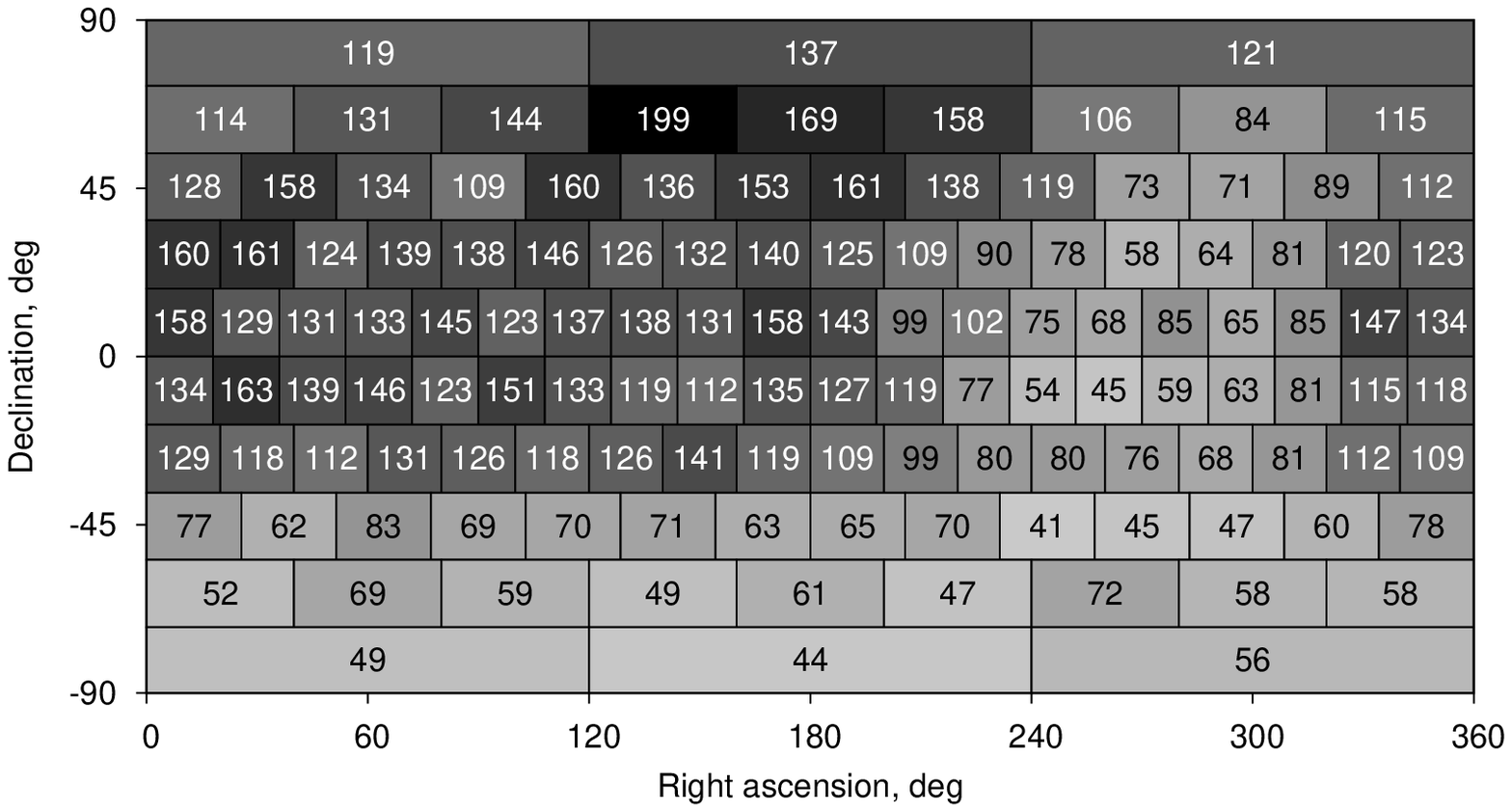}
\caption{Sky distribution of sources for ICRF3-SX (top, 4536 sources) and OCARS (bottom, 13560 sources).}
\label{fig:sky_all_sources}
\end{figure}

Figure~\ref{fig:sky_all_sources} shows the sky distribution of the ICRF3-SX and OCARS sources.
The upper panel of this plot allows us not only to see the well-known weakness of the ICRF3 catalog in the south,
but also to identify the sky regions where the ICRF list should be enriched with new sources to obtain more even
source distribution over the sky, in particular, the region of the Galactic plane.
It can be seen that the number of sources in the poorly populated cells in the OCARS catalog is two-four times larger
than in the ICRF3-SX catalog, which opens a new opportunity to select new ICRF sources from the OCARS source list.

\begin{figure}
\centering
\includegraphics[width=0.48\textwidth]{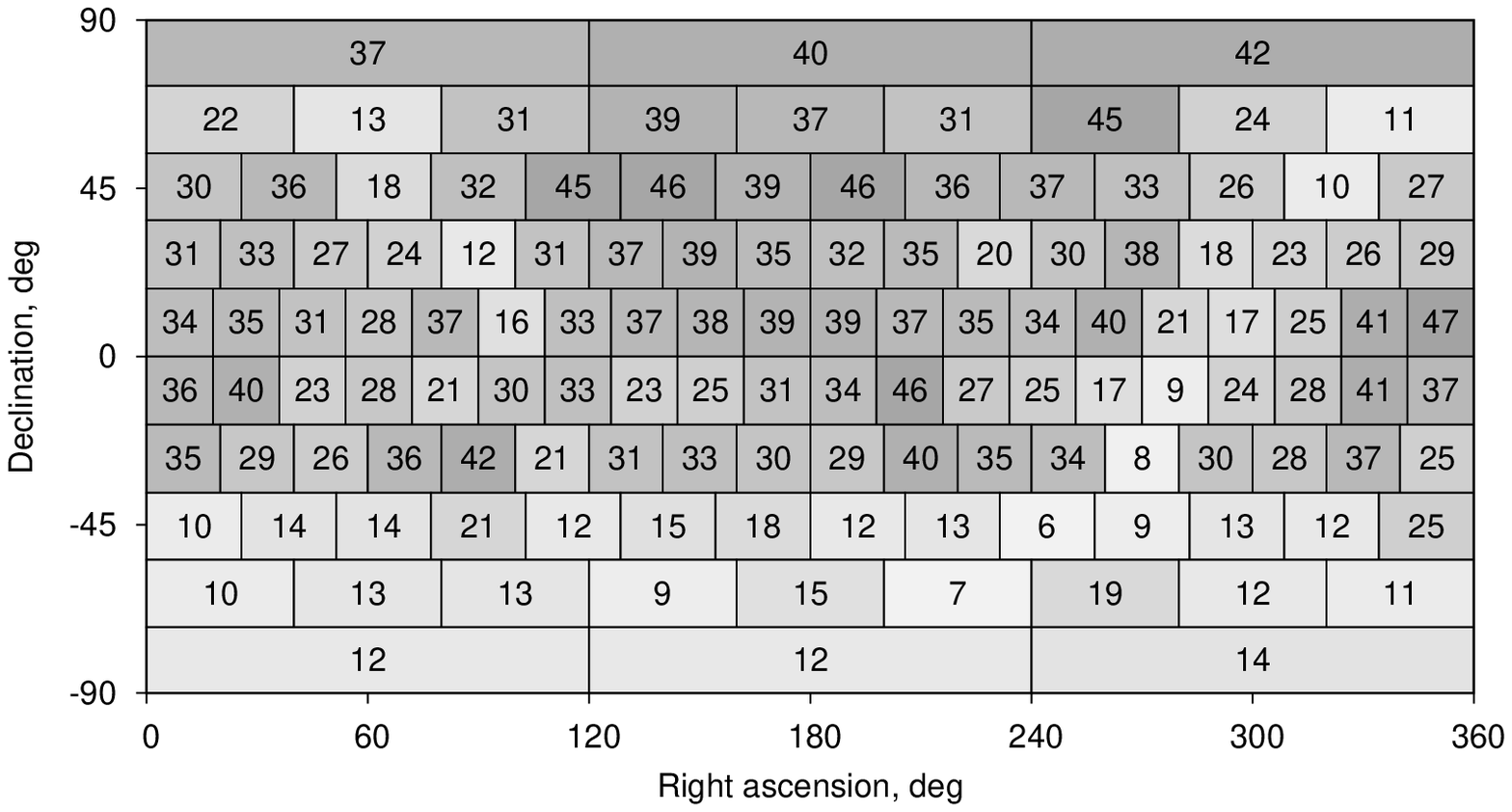}
\includegraphics[width=0.48\textwidth]{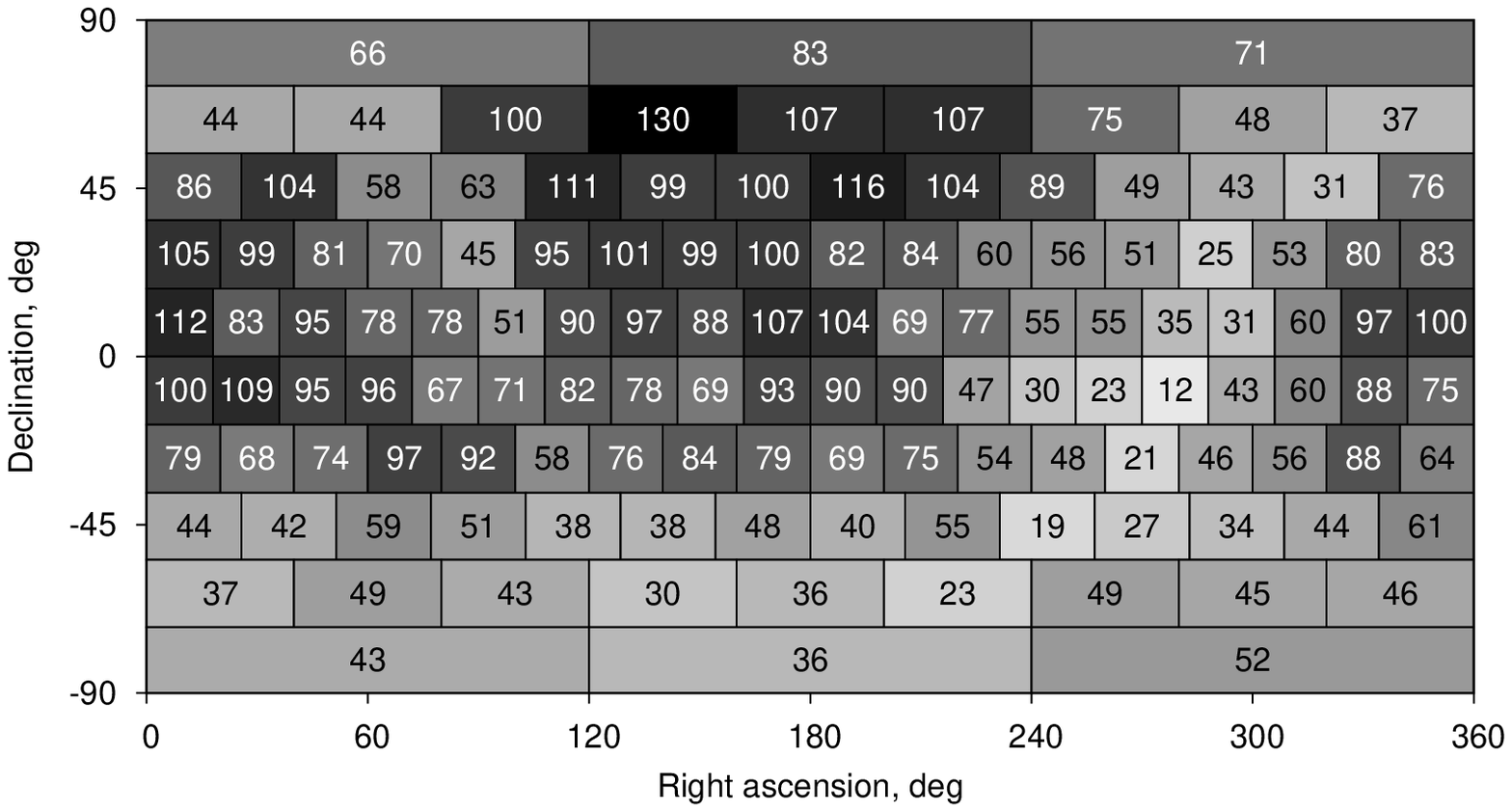}
\caption{Sky distribution of sources cross-identified with $Gaia$ EDR3 for ICRF3-SX (top, 3485 sources)
  and OCARS (bottom, 8707 sources) catalogs.}
\label{fig:sky_gaia}
\end{figure}

Similar situation can be observed when analyzing the sky distribution of common ICRF3-SX and $Gaia$ sources
(Fig.~\ref{fig:sky_gaia}).
In this work, the $Gaia$ EDR3 astrometric catalog \citep{Lindegren2020} was used.
One can see again the source deficiency in the south and near the Galactic plane.
The number of cross-identified sources between the $Gaia$ EDR3 and OCARS catalogs is more than twice as much as
the number of cross-identified sources between the $Gaia$ EDR3 and ICRF3-SX catalogs.
Therefore, the selection of new ICRF sources from the list of cross-identified $Gaia$/OCARS sources will allow us
to improve the uniformity of the source distribution of common ICRF/$Gaia$ sources and thus improve the accuracy,
both systematic and stochastic, of the ICRF/$Gaia$ link.
It should be noted that uneven distribution of common sources may cause problems with analysis of the coordinate
differences ICRF minus $Gaia$ by means of spherical harmonics.

\begin{figure}
\centering
\includegraphics[width=0.48\textwidth]{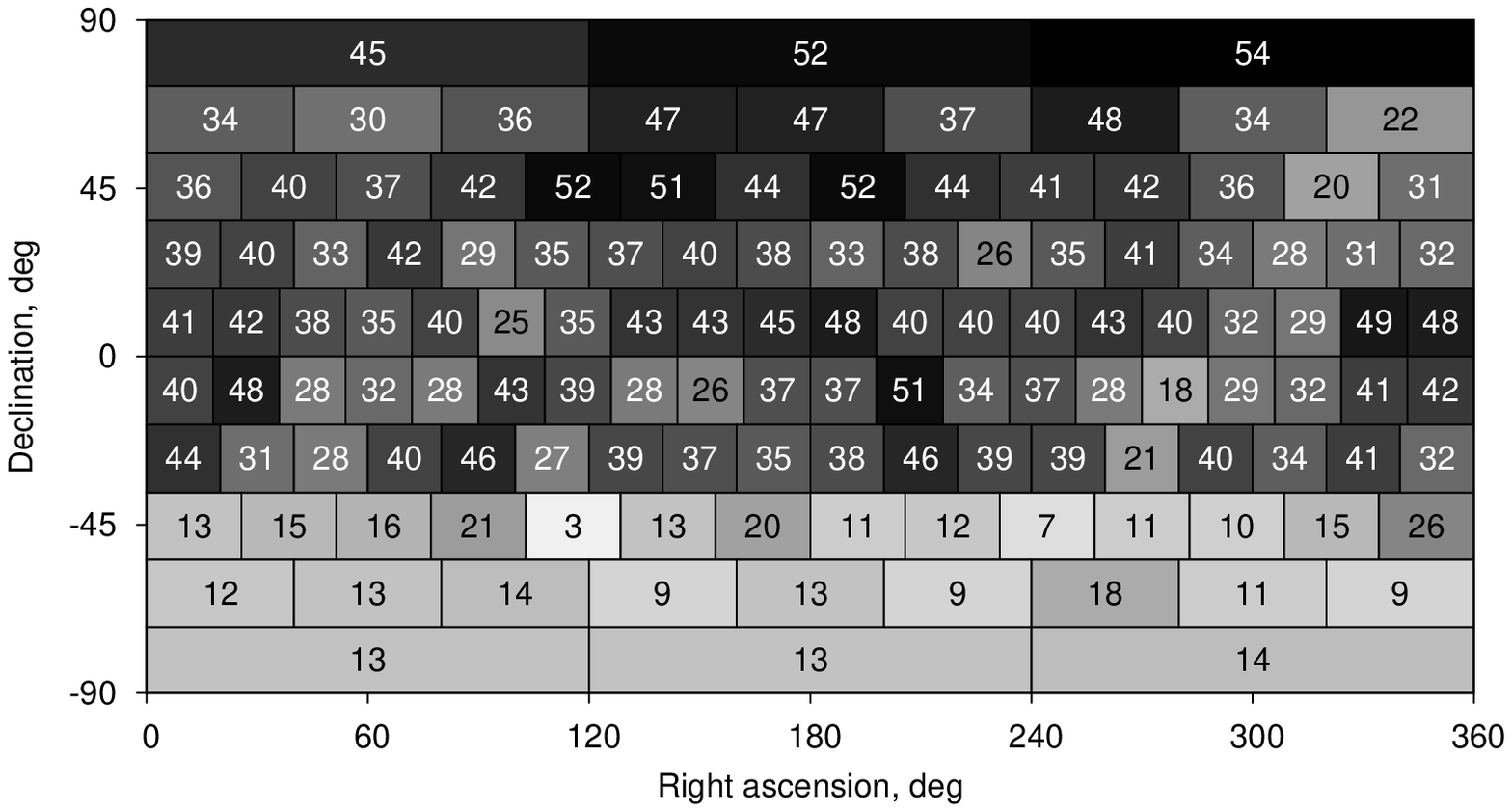}
\includegraphics[width=0.48\textwidth]{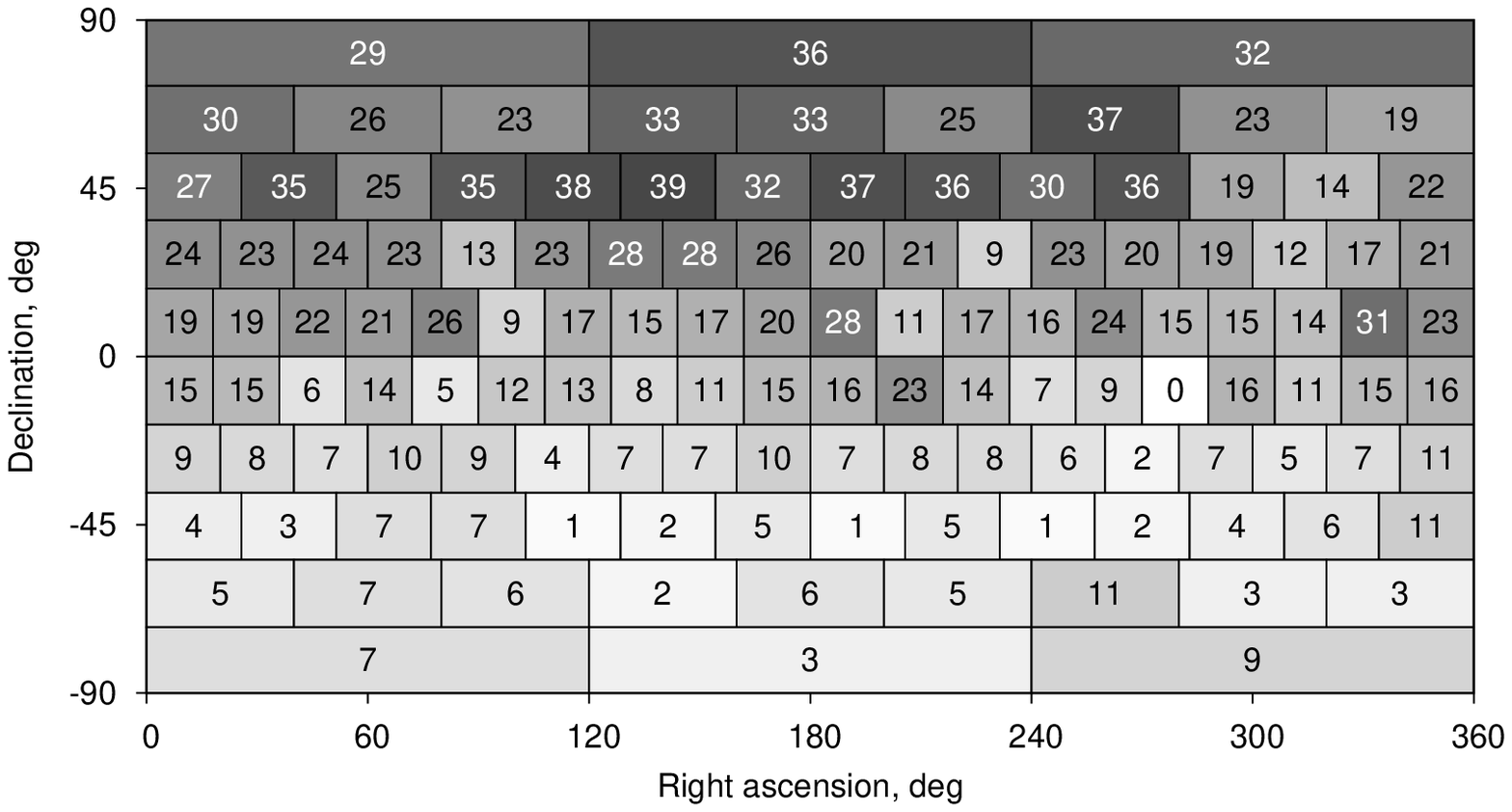}
\caption{Sky distribution of the ICRF3-SX sources with the position errors less than 1~mas (top, 4177 sources)
  and less than 0.2~mas (bottom, 2023 sources).}
\label{fig:sky_errors}
\end{figure}

It is important for ICRF not only to have even source distribution over the sky but also even sky distribution
of the position error.
Figure~\ref{fig:sky_errors} shows the sky distribution of the ICRF3-SX sources with the position errors
less than 1~mas and less than 0.2~mas.
For this work, the source position error is computed as the semi-major axis of the error ellipse.
This analysis allows us to identify the sky regions were additional observations of ICRF sources
should be primarily planned.

\begin{figure}
\centering
\includegraphics[width=0.48\textwidth]{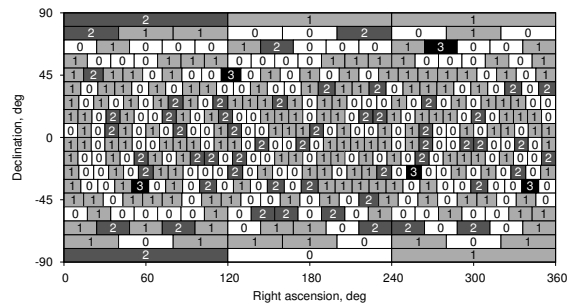}
\caption{Sky distribution of the ICRF3 defining sources.}
\label{fig:sky_defining}
\end{figure}

An important characteristics of the ICRF catalog is the number of defining sources and
how evenly they are distributed over the sky.
The nearest goal may be to have $\approx$400 ICRF4 defining sources (one per 100~sq.~deg).
Figure~\ref{fig:sky_defining} shows the sky distribution of the ICRF3 defining sources.
One can see that there are many cells that lack defining sources.
So, the future ICRF release should take this into account.
Similar analysis was made earlier by \citet{Basu2018}. 

Another consideration which is worth bearing in mind is having more ICRF sources with high redshift.
These sources are important for astrophysical and cosmological studies \cite{Coppejans2017}.
Therefore this makes sense to include the redshift as a criterion for source selection for the ICRF extension.
OCARS provides redshift info for about 58\% of the sources.
Table~\ref{tab:redshift} shows the number of sources with high redshift in the ICRF3 and OCARS catalogs.
The number of high-z sources in the OCARS catalog is about two-four times larger that in the ICRF3 catalog.
So, inclusion of additional high-z sources from OCARS will make the future ICRF releases more useful for
non-astrometric astronomical studies.

\begin{table}
\centering
\caption{Number of sources with high redshift in the ICRF3 and OCARS catalogs.}
\tabcolsep 10pt
\begin{tabular}{lcccc}
\hline
Catalog & \multicolumn{4}{c}{Redshift} \\
& all & $\le 3$ & $\le 4$ & $\le 5$ \\
\hline
ICRF3 & 3382 & 106 & 15 & 3 \\
OCARS & 7967 & 213 & 58 & 13 \\
\hline
\end{tabular}
\label{tab:redshift}
\end{table}

Increasing the number of observations of prospective ICRF sources requires either to involve supplement network
resources, which is not always possible, or to improve the scheduling strategy.
A possible way to make more observations of CRF sources was discussed and tested in \citet{Malkin2013}.

We started with actual schedule for IVS session R1591 that involved the 11-station network.
In the original (actual) IVS schedule for the R1541 session, 60 sources were observed including 7 southern sources
with declination less than $-40^{\circ}$.
For comparisons, the supplementary southern sources were added to the original source list and three experimental schedules
were obtained to evaluate the trade-off between the number of southern sources and the accuracy of geodetic products.
Schedule `R1' was obtained with the original R1591 source list.
Schedule `R1+' includes three more southern sources, and schedule `R1++' includes six more southern sources as compared
with the original R1541 schedule.
The three schedules for 24-hour continuous observations were generated with VieVS scheduling package (Sun et al., 2011).

For Monte Carlo simulation, 50 sessions were generated using the same 24-hour schedule but different realizations of noise delays,
each time creating new values for wet zenith delay, clocks and white noise to simulate observations as realistic as possible.
The simulated NGS data files were entered into the software package VieVS (B\"ohm et al. 2012), which computes a classical least squares solution.
The source coordinates were fixed to the ICRF2 positions, and only Earth orientation parameters (EOP) and station positions were estimated.
The standard deviation of the 50 EOP estimates and mean formal uncertainties obtained in our computations are listed in Table~\ref{tab:param-err}.
One can see that we found no overall degradation of the EOP accuracy after the inclusion of supplement southern sources.
Errors in some EOP became even smaller with inclusion of more southern sources, and some EOP showed minor degradation in the accuracy.

\begin{table}
\centering
\caption{Repeatability and standard deviation of EOP for the IVS R1541 and two experimental schedules R1+ and R1++ (Malkin et al. 2013).}
\label{tab:param-err}
\tabcolsep=6pt
\begin{tabular}{lcccc}
\hline
\multicolumn{1}{c}{Parameter} &     &  R1   &  R1+  &  R1++ \\
\hline
Number of scans               &     &  1258 &  1351 &  1375 \\
Number of observations        &     &  3905 &  3813 &  3997 \\
\hline
EOP repeatability             & Xp  & 143.2 & 125.5 & ~98.2 \\
{[$\mu$as, $\mu$s]}           & Yp  & ~98.2 & ~79.1 & ~96.8 \\
                              & UT1 & ~~5.6 & ~~4.6 & ~~5.9 \\
                              & dX  & ~36.2 & ~42.8 & ~39.1 \\
                              & dY  & ~45.0 & ~39.5 & ~37.2 \\
\hline
Mean EOP uncertainty          & Xp  & ~94.8 & ~95.6 & ~93.4 \\
{[$\mu$as, $\mu$s]}           & Yp  & ~77.2 & ~77.3 & ~74.8 \\
                              & UT1 & ~~4.4 & ~~4.6 & ~~4.7 \\
                              & dX  & ~29.8 & ~30.9 & ~29.5 \\
                              & dY  & ~29.1 & ~29.6 & ~28.1 \\
\hline
\end{tabular}
\end{table}

As to the baseline length repeatability, it was found that for the baselines shorter than $\sim$5,000~km
the R1 schedule shows the best result, and R1+ and R1++ schedules shows worse repeatability, whereas for longer
baselines the R1++ schedule is the best, and R1 is the worst.
However, in fact, the results obtained with the three schedules are close to each other.
The mean baseline length repeatability derived from R1, R1+, and R1++ schedules are 13.5 mm, 12.4 mm, and 11.9 mm, respectively.
In other words, increasing of the number of southern sources (cf. R++ and R+ schedules) leads to a small degradation
of baseline length repeatability for short baselines, and small improvement for long baselines.
However, an overall improvement in the baseline length repeatability was found after inclusion more southern sources in the schedule.

Summarizing the results of this simulation experiment, one can conclude that the inclusion of targeted southern sources
in regular IVS sessions such as R1 and R4 can help to increase substantially the number of ICRF southern sources
with accurate positions.
Suppose, we want to add 100 new sources in the south observed each 100 times (to obtain sub-mas position error) during one year. 
Then we need to make $\sim$200 observations of these sources per week (cf. current $\sim$10K observations in R1+R4 weekly schedule),
which would not significantly influence the normal IVS operations, and might even provide some improvement in obtained geodetic parameters.

\section{Conclusion}

In this paper, an approach is discussed to select new sources for ICRF extension and identify sources already
included in the ICRF that preferably need more observations to improve their position.
It is suggested that the OCARS catalog \citep{Malkin2018}) can be used as an initial (candidate) list of already
VLBI-detected sources to enrich ICRF.

Of course, some sources selected from OCARS using sky coverage criterion discussed in this paper may be too weak
and/or may have a bad structure index (SI) to be used in the regular geodetic observing programs aimed at EOP
and TRF monitoring.
However, both source SI and flux are often variable and change with time \citep{Charlot2008,Charlot2010}.
So, currently ``bad'' source can become a ``good'' one and vice versa.

It should be emphasized that the ICRF is not only needed for geodesy.
Many other scientific and practical applications, from navigation to astrophysics and cosmology, will
appreciate a dense and highly accurate catalog of radio source positions evenly distributed over the sky.
Low flux radio sources are also important for astronomy, especially those sources that are cross-identified
with objects in other wave bands such as optics, X-ray or gamma-ray.
Such cross-identification is also provided by OCARS.

It should be also noted that increasing the number of ICRF sources evenly distributed over the sky is important
for improving the link between ICRF and $Gaia$-CRF.
Therefore, inclusion of radio weak but optically bright sources in ICRF will be of mutual benefit.
OCARS provides optical and NIR magnitudes for about 78\% of the sources, which can be used for source selection.
The selection of candidate sources for ICRF extension that can be used for the ICRF--$Gaia$ link can be made
even simpler because OCARS provides the cross-identification table with the $Gaia$ catalog.

Therefore it is worth spending efforts to enrich the next ICRF catalogs not only with sources ``convenient''
for solving say IERS- and GGOS-related tasks, but also with other sources that would improve general ICRF quality. 
In particular, observations of weak radio sources for ICRF can be performed in cooperation with other VLBI networks,
which include large antennas.

\section{Acknowledgments}

This research has made use of the SAO/NASA Astrophysics Data System\footnote{https://ui.adsabs.harvard.edu/} (ADS).
The figures were prepared using \verb"gnuplot"\footnote{http://www.gnuplot.info/}.


\begin{thebibliography}{99}

\providecommand{\natexlab}[1]{#1}
\providecommand{\url}[1]{\texttt{#1}}
\expandafter\ifx\csname urlstyle\endcsname\relax
  \providecommand{\doi}[1]{doi: #1}\else
  \providecommand{\doi}{doi: \begingroup \urlstyle{rm}\Url}\fi

\bibitem[Basu et al.(2018)]{Basu2018}
Basu S, de Witt A, Quick J, Malkin Z (2018)
\newblock Multi-epoch VLBI images to study the ICRF-3 Defining Sources in the Southern Hemisphere.
\newblock In: \emph{14th European VLBI Network Symposium \& Users Meeting. 8-11 October. Granada, Spain}, PoS(EVN2018)135,
\newblock \doi{10.22323/1.344.0135}

\bibitem[Charlot(2008)]{Charlot2008}
Charlot~P (2008)
\newblock Astrophysical Stability of Radio Sources and Implication for the Realization of the Next ICRF.
\newblock In: A.~Finkelstein and D.~Behrend (eds.) \emph{5th IVS General Meeting Proceedings, St.~Petersburg, Russia, 2008}, 345--354

\bibitem[Charlot et~al.(2010)]{Charlot2010}
Charlot~P, Boboltz~DA, Fey~AL, \textit{et~al.} (2010)
\newblock The celestial reference frame at 24 and 43 GHz. II. Imaging.
\newblock \emph{AJ}, 139, 1713,
\newblock \doi{10.1088/0004-6256/139/5/1713}

\bibitem[Charlot et~al.(2020)]{Charlot2020}
Charlot~P, Jacobs~CS, Gordon~D, \textit{et~al.} (2020)
\newblock The third realization of the International Celestial Reference Frame by very long baseline interferometry.
\newblock \emph{A\&A}, 644, 159,
\newblock \doi{10.1051/0004-6361/202038368}

\bibitem[Coppejans et~al.(2017)]{Coppejans2017}
Coppejans~R, van~Velzen~S; Intema~HT, \textit{et~al.} (2020)
\newblock Radio spectra of bright compact sources at z $>$ 4.5.
\newblock \emph{MNRAS}, 467, 2039
\newblock \doi{10.1093/mnras/stx215}

\bibitem[Lindegren et al.(2020)]{Lindegren2020}
Lindegren L, Klioner SA, Hernandez J, \textit{et al.} (2020)
\newblock $Gaia$ Early Data Release 3: The astrometric solution.
\newblock \emph{A\&A},
\newblock \doi{10.1051/0004-6361/202039709}

\bibitem[Malkin et al.(2013)]{Malkin2013}
Malkin Z, Sun J, B\"ohm J, B\"ohm~S, Kr\'asn\'a~H (2013)
\newblock Searching for an Optimal Strategy to Intensify Observations of the Southern ICRF sources in the framework
 of the regular IVS observing programs.
\newblock In: N.~Zubko and M.~Poutanen (eds.) \emph{21st Meeting of the European VLBI Group for Geodesy and Astronomy, Espoo, Finland, March 5--8, 2013}, 199--204

\bibitem[Malkin(2018)]{Malkin2018}
Malkin~Z (2020)
\newblock A New Version of the OCARS Catalog of Optical Characteristics of Astrometric Radio Sources.
\newblock \emph{AJ}, 239, 20,
\newblock \doi{10.3847/1538-4365/aae777}

\bibitem[Malkin(2019)]{Malkin2019}
Malkin~Z (2019)
\newblock A New Equal-area Isolatitudinal Grid on a Spherical Surface.
\newblock \emph{AJ}, 158, 158,
\newblock \doi{10.3847/1538-3881/ab3a44}

\bibitem[Malkin(2020)]{Malkin2020}
Malkin~Z (2020)
\newblock Spherical Rectangular Equal-Area Grid (SREAG)--Some features.
\newblock In: C.~Bizouard (ed.) \emph{Proc. Journees 2019 Astrometry, Earth Rotation, and Reference Systems in the GAIA era}, 55--59

\end{thebibliography}
\end{document}